%% file: paper.tex
\documentclass[letterpaper,compsoc,twoside]{IEEEtran}
\usepackage{fixltx2e} 
\usepackage{cmap} 
\usepackage{ifthen}
\usepackage[T1]{fontenc}
\usepackage[utf8]{inputenc}
\usepackage{amsmath}

\usepackage[font={small,it},labelfont=bf]{caption}
\usepackage{float}

\usepackage{longtable,ltcaption,array}
\newlength{\DUtablewidth} 

\pdfoutput=1
\usepackage{scipy}
\makeatletter
\def\PY@reset{\let\PY@it=\relax \let\PY@bf=\relax%
    \let\PY@ul=\relax \let\PY@tc=\relax%
    \let\PY@bc=\relax \let\PY@ff=\relax}
\def\PY@tok#1{\csname PY@tok@#1\endcsname}
\def\PY@toks#1+{\ifx\relax#1\empty\else%
    \PY@tok{#1}\expandafter\PY@toks\fi}
\def\PY@do#1{\PY@bc{\PY@tc{\PY@ul{%
    \PY@it{\PY@bf{\PY@ff{#1}}}}}}}
\def\PY#1#2{\PY@reset\PY@toks#1+\relax+\PY@do{#2}}

\expandafter\def\csname PY@tok@gd\endcsname{\def\PY@tc##1{\textcolor[rgb]{0.63,0.00,0.00}{##1}}}
\expandafter\def\csname PY@tok@gu\endcsname{\let\PY@bf=\textbf\def\PY@tc##1{\textcolor[rgb]{0.50,0.00,0.50}{##1}}}
\expandafter\def\csname PY@tok@gt\endcsname{\def\PY@tc##1{\textcolor[rgb]{0.00,0.27,0.87}{##1}}}
\expandafter\def\csname PY@tok@gs\endcsname{\let\PY@bf=\textbf}
\expandafter\def\csname PY@tok@gr\endcsname{\def\PY@tc##1{\textcolor[rgb]{1.00,0.00,0.00}{##1}}}
\expandafter\def\csname PY@tok@cm\endcsname{\let\PY@it=\textit\def\PY@tc##1{\textcolor[rgb]{0.25,0.50,0.56}{##1}}}
\expandafter\def\csname PY@tok@vg\endcsname{\def\PY@tc##1{\textcolor[rgb]{0.73,0.38,0.84}{##1}}}
\expandafter\def\csname PY@tok@m\endcsname{\def\PY@tc##1{\textcolor[rgb]{0.13,0.50,0.31}{##1}}}
\expandafter\def\csname PY@tok@mh\endcsname{\def\PY@tc##1{\textcolor[rgb]{0.13,0.50,0.31}{##1}}}
\expandafter\def\csname PY@tok@cs\endcsname{\def\PY@tc##1{\textcolor[rgb]{0.25,0.50,0.56}{##1}}\def\PY@bc##1{\setlength{\fboxsep}{0pt}\colorbox[rgb]{1.00,0.94,0.94}{\strut ##1}}}
\expandafter\def\csname PY@tok@ge\endcsname{\let\PY@it=\textit}
\expandafter\def\csname PY@tok@vc\endcsname{\def\PY@tc##1{\textcolor[rgb]{0.73,0.38,0.84}{##1}}}
\expandafter\def\csname PY@tok@il\endcsname{\def\PY@tc##1{\textcolor[rgb]{0.13,0.50,0.31}{##1}}}
\expandafter\def\csname PY@tok@go\endcsname{\def\PY@tc##1{\textcolor[rgb]{0.20,0.20,0.20}{##1}}}
\expandafter\def\csname PY@tok@cp\endcsname{\def\PY@tc##1{\textcolor[rgb]{0.00,0.44,0.13}{##1}}}
\expandafter\def\csname PY@tok@gi\endcsname{\def\PY@tc##1{\textcolor[rgb]{0.00,0.63,0.00}{##1}}}
\expandafter\def\csname PY@tok@gh\endcsname{\let\PY@bf=\textbf\def\PY@tc##1{\textcolor[rgb]{0.00,0.00,0.50}{##1}}}
\expandafter\def\csname PY@tok@ni\endcsname{\let\PY@bf=\textbf\def\PY@tc##1{\textcolor[rgb]{0.84,0.33,0.22}{##1}}}
\expandafter\def\csname PY@tok@nl\endcsname{\let\PY@bf=\textbf\def\PY@tc##1{\textcolor[rgb]{0.00,0.13,0.44}{##1}}}
\expandafter\def\csname PY@tok@nn\endcsname{\let\PY@bf=\textbf\def\PY@tc##1{\textcolor[rgb]{0.05,0.52,0.71}{##1}}}
\expandafter\def\csname PY@tok@no\endcsname{\def\PY@tc##1{\textcolor[rgb]{0.38,0.68,0.84}{##1}}}
\expandafter\def\csname PY@tok@na\endcsname{\def\PY@tc##1{\textcolor[rgb]{0.25,0.44,0.63}{##1}}}
\expandafter\def\csname PY@tok@nb\endcsname{\def\PY@tc##1{\textcolor[rgb]{0.00,0.44,0.13}{##1}}}
\expandafter\def\csname PY@tok@nc\endcsname{\let\PY@bf=\textbf\def\PY@tc##1{\textcolor[rgb]{0.05,0.52,0.71}{##1}}}
\expandafter\def\csname PY@tok@nd\endcsname{\let\PY@bf=\textbf\def\PY@tc##1{\textcolor[rgb]{0.33,0.33,0.33}{##1}}}
\expandafter\def\csname PY@tok@ne\endcsname{\def\PY@tc##1{\textcolor[rgb]{0.00,0.44,0.13}{##1}}}
\expandafter\def\csname PY@tok@nf\endcsname{\def\PY@tc##1{\textcolor[rgb]{0.02,0.16,0.49}{##1}}}
\expandafter\def\csname PY@tok@si\endcsname{\let\PY@it=\textit\def\PY@tc##1{\textcolor[rgb]{0.44,0.63,0.82}{##1}}}
\expandafter\def\csname PY@tok@s2\endcsname{\def\PY@tc##1{\textcolor[rgb]{0.25,0.44,0.63}{##1}}}
\expandafter\def\csname PY@tok@vi\endcsname{\def\PY@tc##1{\textcolor[rgb]{0.73,0.38,0.84}{##1}}}
\expandafter\def\csname PY@tok@nt\endcsname{\let\PY@bf=\textbf\def\PY@tc##1{\textcolor[rgb]{0.02,0.16,0.45}{##1}}}
\expandafter\def\csname PY@tok@nv\endcsname{\def\PY@tc##1{\textcolor[rgb]{0.73,0.38,0.84}{##1}}}
\expandafter\def\csname PY@tok@s1\endcsname{\def\PY@tc##1{\textcolor[rgb]{0.25,0.44,0.63}{##1}}}
\expandafter\def\csname PY@tok@gp\endcsname{\let\PY@bf=\textbf\def\PY@tc##1{\textcolor[rgb]{0.78,0.36,0.04}{##1}}}
\expandafter\def\csname PY@tok@sh\endcsname{\def\PY@tc##1{\textcolor[rgb]{0.25,0.44,0.63}{##1}}}
\expandafter\def\csname PY@tok@ow\endcsname{\let\PY@bf=\textbf\def\PY@tc##1{\textcolor[rgb]{0.00,0.44,0.13}{##1}}}
\expandafter\def\csname PY@tok@sx\endcsname{\def\PY@tc##1{\textcolor[rgb]{0.78,0.36,0.04}{##1}}}
\expandafter\def\csname PY@tok@bp\endcsname{\def\PY@tc##1{\textcolor[rgb]{0.00,0.44,0.13}{##1}}}
\expandafter\def\csname PY@tok@c1\endcsname{\let\PY@it=\textit\def\PY@tc##1{\textcolor[rgb]{0.25,0.50,0.56}{##1}}}
\expandafter\def\csname PY@tok@kc\endcsname{\let\PY@bf=\textbf\def\PY@tc##1{\textcolor[rgb]{0.00,0.44,0.13}{##1}}}
\expandafter\def\csname PY@tok@c\endcsname{\let\PY@it=\textit\def\PY@tc##1{\textcolor[rgb]{0.25,0.50,0.56}{##1}}}
\expandafter\def\csname PY@tok@mf\endcsname{\def\PY@tc##1{\textcolor[rgb]{0.13,0.50,0.31}{##1}}}
\expandafter\def\csname PY@tok@err\endcsname{\def\PY@bc##1{\setlength{\fboxsep}{0pt}\fcolorbox[rgb]{1.00,0.00,0.00}{1,1,1}{\strut ##1}}}
\expandafter\def\csname PY@tok@kd\endcsname{\let\PY@bf=\textbf\def\PY@tc##1{\textcolor[rgb]{0.00,0.44,0.13}{##1}}}
\expandafter\def\csname PY@tok@ss\endcsname{\def\PY@tc##1{\textcolor[rgb]{0.32,0.47,0.09}{##1}}}
\expandafter\def\csname PY@tok@sr\endcsname{\def\PY@tc##1{\textcolor[rgb]{0.14,0.33,0.53}{##1}}}
\expandafter\def\csname PY@tok@mo\endcsname{\def\PY@tc##1{\textcolor[rgb]{0.13,0.50,0.31}{##1}}}
\expandafter\def\csname PY@tok@mi\endcsname{\def\PY@tc##1{\textcolor[rgb]{0.13,0.50,0.31}{##1}}}
\expandafter\def\csname PY@tok@kn\endcsname{\let\PY@bf=\textbf\def\PY@tc##1{\textcolor[rgb]{0.00,0.44,0.13}{##1}}}
\expandafter\def\csname PY@tok@o\endcsname{\def\PY@tc##1{\textcolor[rgb]{0.40,0.40,0.40}{##1}}}
\expandafter\def\csname PY@tok@kr\endcsname{\let\PY@bf=\textbf\def\PY@tc##1{\textcolor[rgb]{0.00,0.44,0.13}{##1}}}
\expandafter\def\csname PY@tok@s\endcsname{\def\PY@tc##1{\textcolor[rgb]{0.25,0.44,0.63}{##1}}}
\expandafter\def\csname PY@tok@kp\endcsname{\def\PY@tc##1{\textcolor[rgb]{0.00,0.44,0.13}{##1}}}
\expandafter\def\csname PY@tok@w\endcsname{\def\PY@tc##1{\textcolor[rgb]{0.73,0.73,0.73}{##1}}}
\expandafter\def\csname PY@tok@kt\endcsname{\def\PY@tc##1{\textcolor[rgb]{0.56,0.13,0.00}{##1}}}
\expandafter\def\csname PY@tok@sc\endcsname{\def\PY@tc##1{\textcolor[rgb]{0.25,0.44,0.63}{##1}}}
\expandafter\def\csname PY@tok@sb\endcsname{\def\PY@tc##1{\textcolor[rgb]{0.25,0.44,0.63}{##1}}}
\expandafter\def\csname PY@tok@k\endcsname{\let\PY@bf=\textbf\def\PY@tc##1{\textcolor[rgb]{0.00,0.44,0.13}{##1}}}
\expandafter\def\csname PY@tok@se\endcsname{\let\PY@bf=\textbf\def\PY@tc##1{\textcolor[rgb]{0.25,0.44,0.63}{##1}}}
\expandafter\def\csname PY@tok@sd\endcsname{\let\PY@it=\textit\def\PY@tc##1{\textcolor[rgb]{0.25,0.44,0.63}{##1}}}

\def\PYZus{\char`\_}
\def\PYZob{\char`\{}
\def\PYZcb{\char`\}}

\def\PYZlt{\char`\<}
\def\PYZgt{\char`\>}
\def\PYZsh{\char`\#}

\def\PYZdl{\char`\$}
\def\PYZhy{\char`\-}
\def\PYZsq{\char`\'}
\def\PYZdq{\char`\"}

\makeatother

\providecommand*{\DUrole}[2]{%
  \ifcsname DUrole#1\endcsname%
    \csname DUrole#1\endcsname{#2}%
  \else
    \ifcsname docutilsrole#1\endcsname%
      \csname docutilsrole#1\endcsname{#2}%
    \else%
      #2%
    \fi%
  \fi%
}

\ifthenelse{\isundefined{\hypersetup}}{
  \usepackage[colorlinks=true,linkcolor=blue,urlcolor=blue]{hyperref}
  \urlstyle{same} 
}{}

\begin{document}
\newcounter{footnotecounter}\title{Using Python to Dive into Signalling Data with CellNOpt and BioServices}\author{Thomas Cokelaer$^{\setcounter{footnotecounter}{1}\fnsymbol{footnotecounter}\setcounter{footnotecounter}{2}\fnsymbol{footnotecounter}}$%
          \setcounter{footnotecounter}{1}\thanks{\fnsymbol{footnotecounter} %
          Corresponding author: \protect\href{mailto:cokelaer@ebi.ac.uk}{cokelaer@ebi.ac.uk}}\setcounter{footnotecounter}{2}\thanks{\fnsymbol{footnotecounter} European Bioinformatics Institute (EMBL-EBI)}, Julio Saez-Rodriguez$^{\setcounter{footnotecounter}{2}\fnsymbol{footnotecounter}}$\thanks{%

          \noindent%
          Copyright\,\copyright\,2014 Thomas Cokelaer et al. This is an open-access article distributed under the terms of the Creative Commons Attribution License, which permits unrestricted use, distribution, and reproduction in any medium, provided the original author and source are credited. http://creativecommons.org/licenses/by/3.0/%
        }}\maketitle
          \renewcommand{\leftmark}{PROC. OF THE 7th EUR. CONF. ON PYTHON IN SCIENCE (EUROSCIPY 2014)}
          \renewcommand{\rightmark}{USING PYTHON TO DIVE INTO SIGNALLING DATA WITH CELLNOPT AND BIOSERVICES}

\InputIfFileExists{page_numbers.tex}{}{}
\newcommand*{\docutilsroleref}{\ref}
\newcommand*{\docutilsrolelabel}{\label}
\AtEndDocument{\cleardoublepage}
\begin{abstract}Systems biology is an inter-disciplinary field that  studies systems of
biological components at different scales, which may be molecules, cells or
entire organism. In particular, systems biology methods are applied to
understand functional deregulations within human cells (e.g., cancers). In
this context, we present several python packages linked to \textbf{CellNOptR} (R
package), which is used to build predictive logic models of signalling
networks by training networks (derived from literature) to signalling
(phospho-proteomic) data. The first package (\textbf{cellnopt.wrapper}) is a
wrapper based on RPY2 that allows a full access to CellNOptR
functionalities within Python. The second one (\textbf{cellnopt.core}) was
designed to ease the manipulation and visualisation of data structures used
in CellNOptR, which was achieved by using Pandas, NetworkX and matplotlib.
Systems biology also makes extensive use of web resources and services. We
will give an overview and status of \textbf{BioServices}, which allows one to
access programmatically to web resources used in life science and how it
can be combined with CellNOptR.\end{abstract}

\section{Context and Introduction%
  \label{context-and-introduction}%
}

Systems biology studies systems of biological components at different scales,
which may be molecules, cells or entire organisms. It is a recent term that
emerged in the 2000s \cite{IDE01}, \cite{KIT02} to describe an inter-disciplinary
research field in biology. In human cells, which will be considered in this
paper, systems biology helps to understand functional deregulations inside the
cells that are induced either by gene mutations (in the  nucleus) or
extracellular signalling. Such deregulations may lead to the apparition of
cancers or other diseases.

Cells are constantly stimulated by extracellular signalling. Receptors on the
cell surface may be activated by those signals thereby triggering a chain of
events inside the cell (signal transduction). These chains of events are also
called \textbf{signalling pathways}. Depending on the response, the cell behaviour
may be altered (shape, gene expression, etc.). These pathways are connected to
dense network of interactions between proteins that propagate the external
signals down to the gene expression level (see Figure \DUrole{ref}{fig:overview}). For
simplicity, relationships between proteins are often considered to be either
activation or inhibition. In addition, protein complexes
may also be formed, which means that several type of proteins may be required
to activate or inhibit another protein. Protein interaction networks are
complex:%
\begin{itemize}

\item 

the number of protein types is large (about 20,000 in human cells)
\item 

signalling pathways are context specific and cell-type specific; there are
about 200 human cell types (e.g., blood, liver)
\item 

proteins may have different dynamic (from a few minutes to several hours).
\end{itemize}

A classical pathway example is the so-called P53 pathway (see Figure
\DUrole{ref}{fig:overview}). In a normal cell the P53 protein is inactive (inhibited by
the MDM2 protein). However, upon DNA damage or other stresses, various pathways
will lead to the dissociation of this P53-MDM2 complex, thereby activating P53.
Consequently, P53 will either prevent further cell growth to allow a DNA repair
or initiate an apoptosis (cell death) to discard the damaged cell. A
deregulation of the P53 pathway would results in an uncontrolled cell
proliferation, such as cancer \cite{HAUPT}.\begin{figure}[]\noindent\makebox[\columnwidth][c]{\includegraphics[scale=0.20]{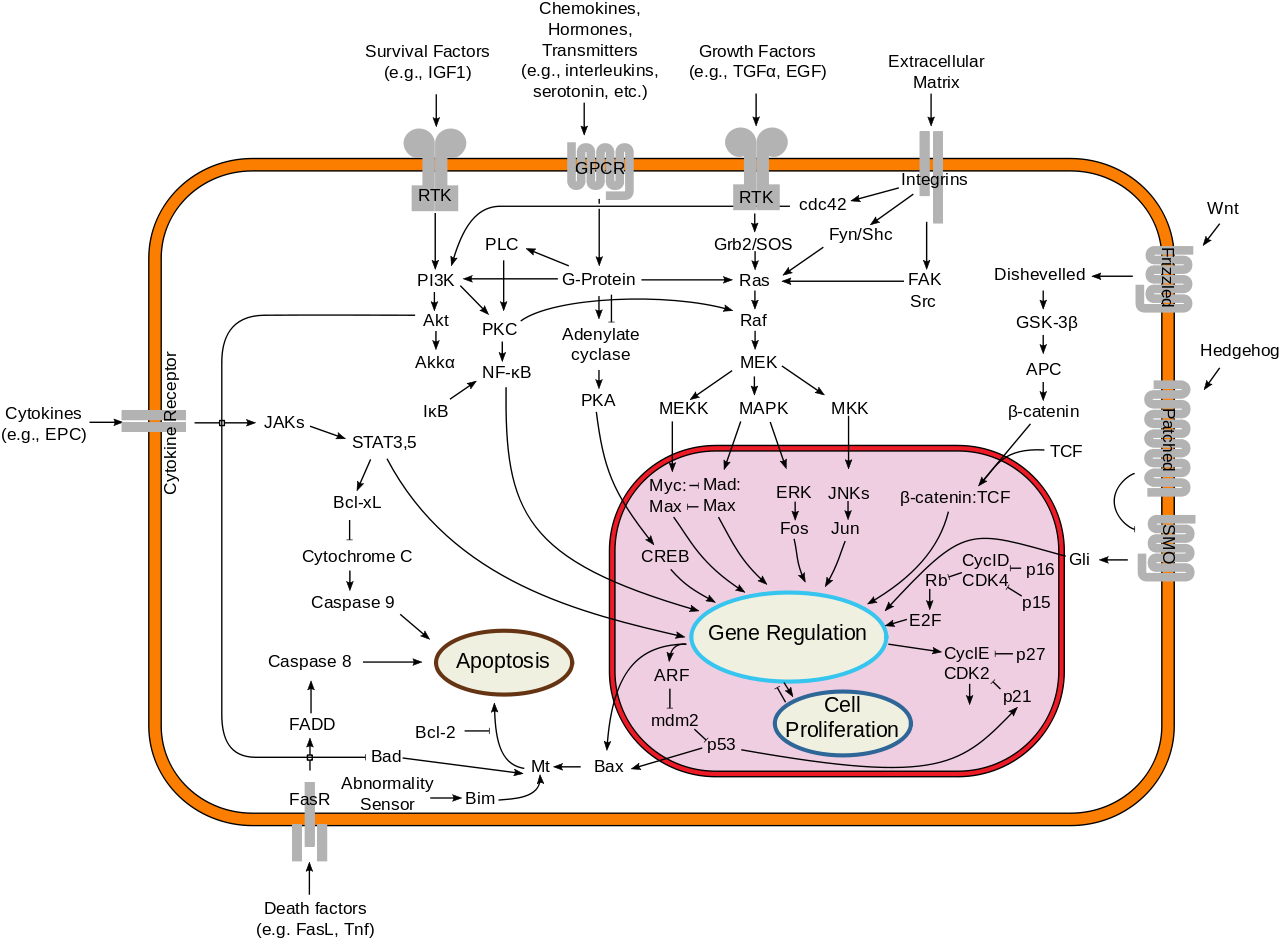}}
\caption{Overview of signal transduction pathways. The cell boundary (thick orange
line) has receptors to external signals (e.g., survival factors IGF1).
These signals propagate inside the cell via complex networks of protein
interactions (e.g., activation and inhibition) down to the gene expression
level inside the cell nucleus (red thick line). Image source: \href{http://en.wikipedia.org/wiki/File:Signal_transduction_v1.png}{wikipedia}.
\DUrole{label}{fig:overview}}
\end{figure}

In order to predict novel therapeutic solutions, it is essential to understand
the behaviour of signalling pathways. Discrete logic modelling provides a
framework to link signalling pathways to extracellular signals and drug effects
\cite{SAEZ}.  Experimental data can be obtained by measuring protein responses to
combination of drugs (altering normal behaviour of a protein) and stimulations.
There are different type of experiments from mass-spectrometry (many proteins
but few perturbations) to antibody-based experiments (few proteins but more time
points and perturbations).

The software CellNOptR \cite{CNO12} provides tools to perform logic modeling at
the protein level using  network of protein interactions and perturbation data
sets. The core of the software consist in (1)
transforming a protein network into a logical network; (2) simulating the flow of
signalling in the network using for instance a boolean formalism; (3) comparing
real biological data with the simulated data. The software is essentially  an
optimisation problem, which can be solved by various algorithms (e.g., genetic
algorithm).

Although CellNOpt is originally written with the R language, we will focus on
two python packages that are related to it. The first one called
\textbf{cellnopt.wrapper} is a Python wrapper that have been written using the RPy2
package. The second package is called \textbf{cellnopt.core}. It combines several
libraries (e.g., Pandas \cite{MCK10}, NetworkX \cite{ARI08} and Matplotlib \cite{HUN07}) to
provide tools dedicated to the manipulation of network
of proteins and perturbation data sets that are the input of CellNOptR packages.

Another important need of systems biology is to be able to access online
resources and databases. In the context of logical modeling, resources of
importance are signalling pathways (e.g., Wiki Pathway \cite{WP09}) and retrieval of
information about proteins (e.g., UniProt \cite{UNI14}). In order to help us in this
task, we developed \textbf{BioServices} \cite{COK13} that ease programmatic access to web services
in Python. It was then extended to retrieve information from other web services
so as to cover the spectrum of bioinformatics resources (e.g., genomics,
sequence analysis).

In the first part of this paper, we will briefly present the data structure
used in CellNOptR and a typical pipeline. We will then demonstrate how
\textbf{cellnopt.wrapper} and \textbf{cellnopt.core} can enhance user experience. In the
second part,  we will quickly present \textbf{BioServices} and give an update on its
status and future directions.

\section{CellNOpt%
  \label{cellnopt}%
}

CellNOptR \cite{CNO12} is a R package used for creating logic-based models of signal
transduction networks using different logic formalisms but we consider boolean
logic only here below. Other formalisms including differential equation
formalism are covered in \cite{MAC12} , \cite{CNO12}.

In a nutshell, CellNOptR uses information on signalling pathways encoded as a
\emph{Prior Knowledge Network (PKN)}, and trains it against high-throughput
\emph{biochemical data} to create cell-specific models. The \emph{training} is performed
with optimisation such as genetic algorithms. For more details see also the
\url{www.cellnopt.org} website.

\subsection{Input data structures%
  \label{input-data-structures}%
}

\subsubsection{Network and logic model%
  \label{network-and-logic-model}%
}
\begin{figure}[]\noindent\makebox[\columnwidth][c]{\includegraphics[scale=0.35]{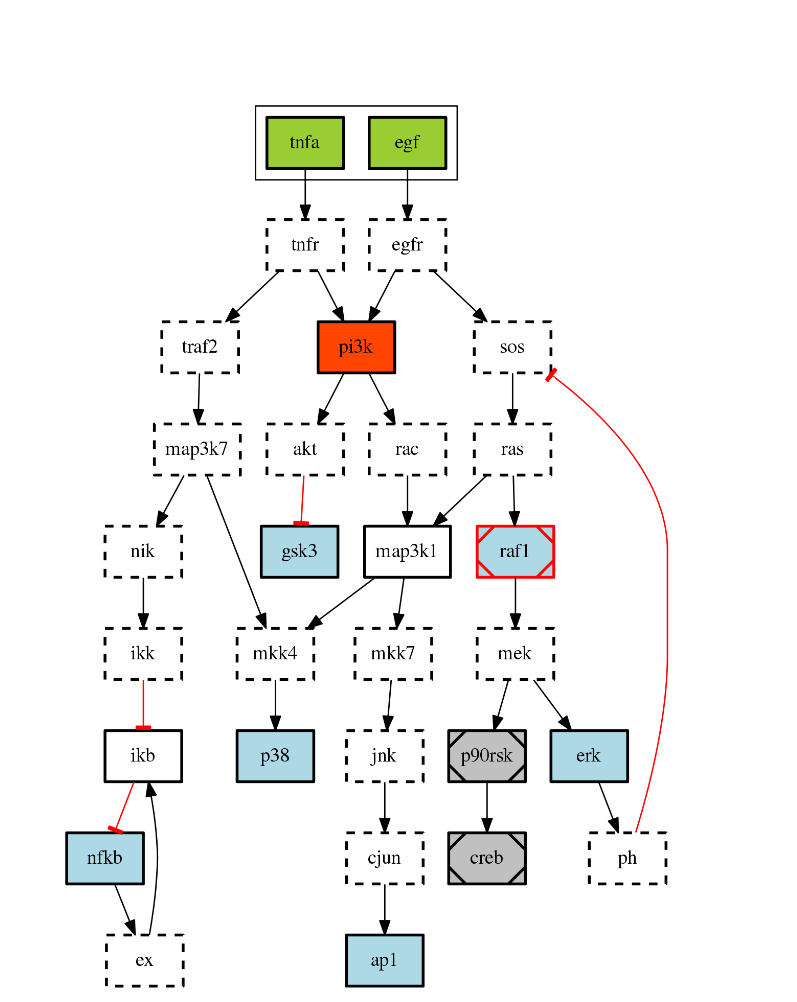}}
\caption{Prior Knowledge Network (PKN) example. Colored nodes represent (i) stimuli
(green, generally on cell surface or close to), (ii) measured proteins
(blue), (iii) inhibited protein by a drug (red), (iv) silent nodes (white
and grey) that do not affect the logic of the model if removed. Black edges
represent activation and red edges represent inhibition. \DUrole{label}{figpkn}}
\end{figure}

The PKNs gives a list of known relationships between proteins. It is built from
literature or expertise from experimentalists.  One way to store the PKNs is to
use  the SIF format, which list relationships between proteins within a
tabulated-separated values file. Consider this example:%
\begin{quote}\begin{verbatim}
Input1 1 Interm
Input2 1 Interm
Interm 1 Output
\end{verbatim}

\end{quote}
Each row is a reaction where the first element is the input protein, the third
element is the affected protein, and the middle  element is the relationship,
where 1 means activation and -1 means inhibition. A visual representation of
this example is shown in Figure \DUrole{ref}{fig:cnoproc}. A more realistic example is
also provided in Figure \DUrole{ref}{figpkn}. Such networks are directed graphs where
edges can be either activation (represented by normal black edge) or inhibition
(represented by the red edge).

In the SIF file provided above, only OR relationships are encoded: the protein
\emph{Interm} is activated by the \emph{Input1} OR \emph{Input2} protein. Within cells, complex
of proteins do exist, which means that an AND relationship is also possible.
Transforming the input PKN into a logical model means that AND gates have to be
added (if there are several inputs).

\subsubsection{Data%
  \label{data}%
}

The data used in CellNOpt is made of measurements of protein responses to
perturbations, which is a combination of stimuli (on cell receptor) and
inhibition (caused by a drug treatment). These measurements are stored in a format
called MIDAS \cite{MIDAS}, which is a CSV file format. Figure \DUrole{ref}{figmidas} gives
an example of a MIDAS data file together with further explanations.

\subsubsection{Training%
  \label{training}%
}

Once a PKN and a MIDAS file are in place, the PKN is transformed into a logic
model. Further simplifications can be applied on the model as shown in Figure
\DUrole{ref}{fig:cnoproc} (e.g., compression to remove nodes/proteins that do not
change the logic of the network). Finally, the training of the logic model to
the data is performed by minimising an objective function written as follows:\begin{equation*}
\theta(M) = \theta_f(M) + \alpha \theta_s(M)
\end{equation*}where\begin{equation*}
\theta_f(M) = \frac{1}{N} \sum_{k=1}^K \sum_{e=1}^E \sum_{t=1}^T  (X_{k,e,t} - X_{k,e,t}^s)^2
\end{equation*}where $e$ is an experiment, $k$ a measured protein and $t$ a
time point. The total number of points is $N=E.K.T$ where E, K and T are
the total number of experiments, measured proteins and time points,
respectively. $X_{k,e,t}$ is a measurement and $X^s_{e,k,t}$ the
corresponding simulated measurement returned by the simulated model $M$. A
model $M$ is a subset of the initial PKN where edges have been pruned (or
not). Finally, $\theta_s$ penalises the model size by summing across the
number of inputs of each edge and $\alpha$ is a tunable parameter.\begin{figure}[]\noindent\makebox[\columnwidth][c]{\includegraphics[width=\columnwidth]{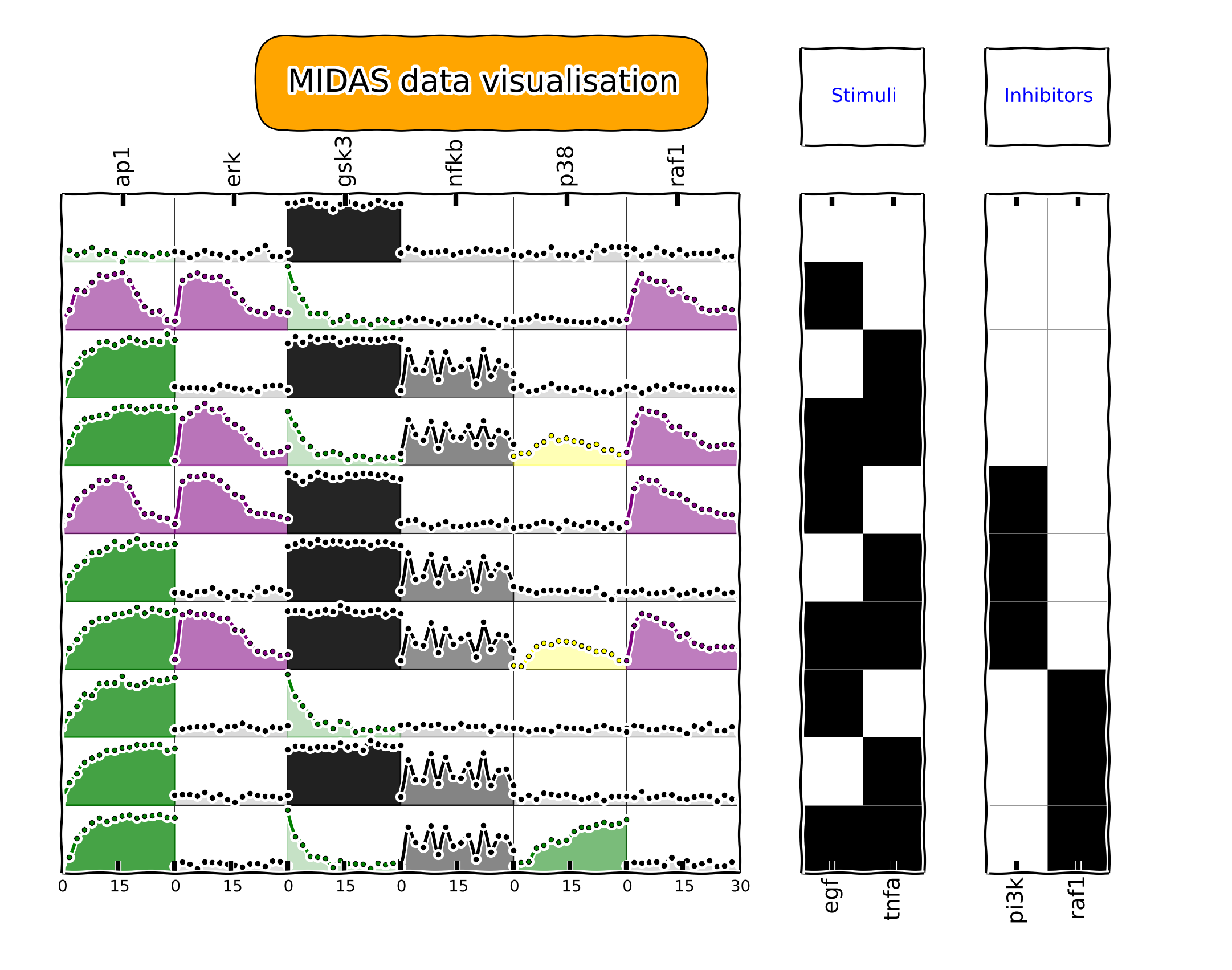}}
\caption{MIDAS data set visualised with cellnopt.core. Each row correspond to an
experiment, that is a combination of stimuli and inhibitors (drug). An
experiment is summarized by the two right panels where the x-axis contains
the name of the stimuli and inhibitors and a black square means stimuli (or
inhibitor) is on. The right panel contains the measurements made on each
protein of interest over time. For example, the left bottom box gives us
about 15 time points for the protein AP1 in the experimental conditions
where EGF and TNFA receptors are stimulated and RAF1 is inhibited. The color
in the boxes indicates the rough trend of the time series (e.g., green
means activation is going up, the alpha transparency indicates the
strength of the signals.). \DUrole{label}{figmidas}}
\end{figure}

\subsection{cellnopt.wrapper%
  \label{cellnopt-wrapper}%
}

CellNOptR provides a set of R packages available on BioConductor website, which
guarantees a minimal quality. Packages are indeed multi-platform and tested
regularly. However, the functional approach that has been chosen limits somehow
the user experience. In order to be able to use the Python language, we
therefore decided to also provide a python wrapper. To do so, we used the
RPY2 package. The cost for the implementation is reasonable: the R
packages in CellNOptR relies on 16,000 lines of code (in R) and another
4,000 in C, while the final python wrappers requires 2000 lines of code
including the documentation.

In addition to the wrappers, we also implemented a set of classes (or for each
of the logical formalism) that encapsulate the R functions. The results is that
\textbf{cellnopt.wrapper} (introduced in \cite{CNO12}) provides a full access to the
entire CellNOptR packages with an objected oriented approach.

A simple R script written with CellNOptR functions (to find the optimal model
that fit the data) would look like:\begin{Verbatim}[commandchars=\\\{\},numbers=left,firstnumber=1,stepnumber=1,fontsize=\footnotesize,xleftmargin=2.25mm,numbersep=3pt]
library\PY{p}{(}CellNOptR\PY{p}{)}
model \PY{o}{=} readSIF\PY{p}{(}CNOdata\PY{p}{(}\PY{l+s}{\PYZdq{}}\PY{l+s}{PKN\PYZhy{}ToyMMB.sif\PYZdq{}}\PY{p}{)}\PY{p}{)}
data \PY{o}{=} CNOlist\PY{p}{(}CNOdata\PY{p}{(}\PY{l+s}{\PYZdq{}}\PY{l+s}{MD\PYZhy{}ToyMMB.csv\PYZdq{}}\PY{p}{)}\PY{p}{)}
res \PY{o}{=} gaBinaryT1\PY{p}{(}data\PY{p}{,} model\PY{p}{)}
plotFit\PY{p}{(}res\PY{p}{)}
cutAndPlotResultsT1\PY{p}{(}model\PY{p}{,} res\PY{o}{\PYZdl{}}bString\PY{p}{,} \PY{k+kc}{NULL}\PY{p}{,} data\PY{p}{)}
\end{Verbatim}
On the first line, we load the library. On the second and third lines, we read
the PKN and MIDAS files. The optimisation is performed with a genetic algorithm
(line 4). We plot the evolution of the objective function over time (line 5) and
finally look at the individual fits (see Figure \DUrole{ref}{figfit} for an example).
Here below is the same code in Python using \textbf{cellnopt.wrapper}\begin{Verbatim}[commandchars=\\\{\},numbers=left,firstnumber=1,stepnumber=1,fontsize=\footnotesize,xleftmargin=2.25mm,numbersep=3pt]
\PY{k+kn}{from} \PY{n+nn}{cellnopt.wrapper} \PY{k+kn}{import} \PY{n}{CNORbool}
\PY{n}{b} \PY{o}{=} \PY{n}{CNORbool}\PY{p}{(}\PY{n}{cnodata}\PY{p}{(}\PY{l+s}{\PYZdq{}}\PY{l+s}{PKN\PYZhy{}ToyMMB.sif}\PY{l+s}{\PYZdq{}}\PY{p}{)}\PY{p}{,}
             \PY{n}{cnodata}\PY{p}{(}\PY{l+s}{\PYZdq{}}\PY{l+s}{MD\PYZhy{}ToyMMB.csv}\PY{l+s}{\PYZdq{}}\PY{p}{)}\PY{p}{)}
\PY{n}{b}\PY{o}{.}\PY{n}{gaBinaryT1}\PY{p}{(}\PY{p}{)}
\PY{n}{b}\PY{o}{.}\PY{n}{plotFit}\PY{p}{(}\PY{p}{)}
\PY{n}{b}\PY{o}{.}\PY{n}{cutAndPlotResultsT1}\PY{p}{(}\PY{p}{)}
\end{Verbatim}
The two code snippets are equivalent. The main difference appears to be that
the first code is functional and the second is object-oriented. The value of
the Python wrapping is that new classes can be derived, introspection of the
data is possible and more importantly further manipulation of the results in
Python is possible. Because an object-oriented approach is used in place of
functional programming, the user interface is also simplified (no need to
provide additional parameters).\begin{figure}[]\noindent\makebox[\columnwidth][c]{\includegraphics[width=\columnwidth]{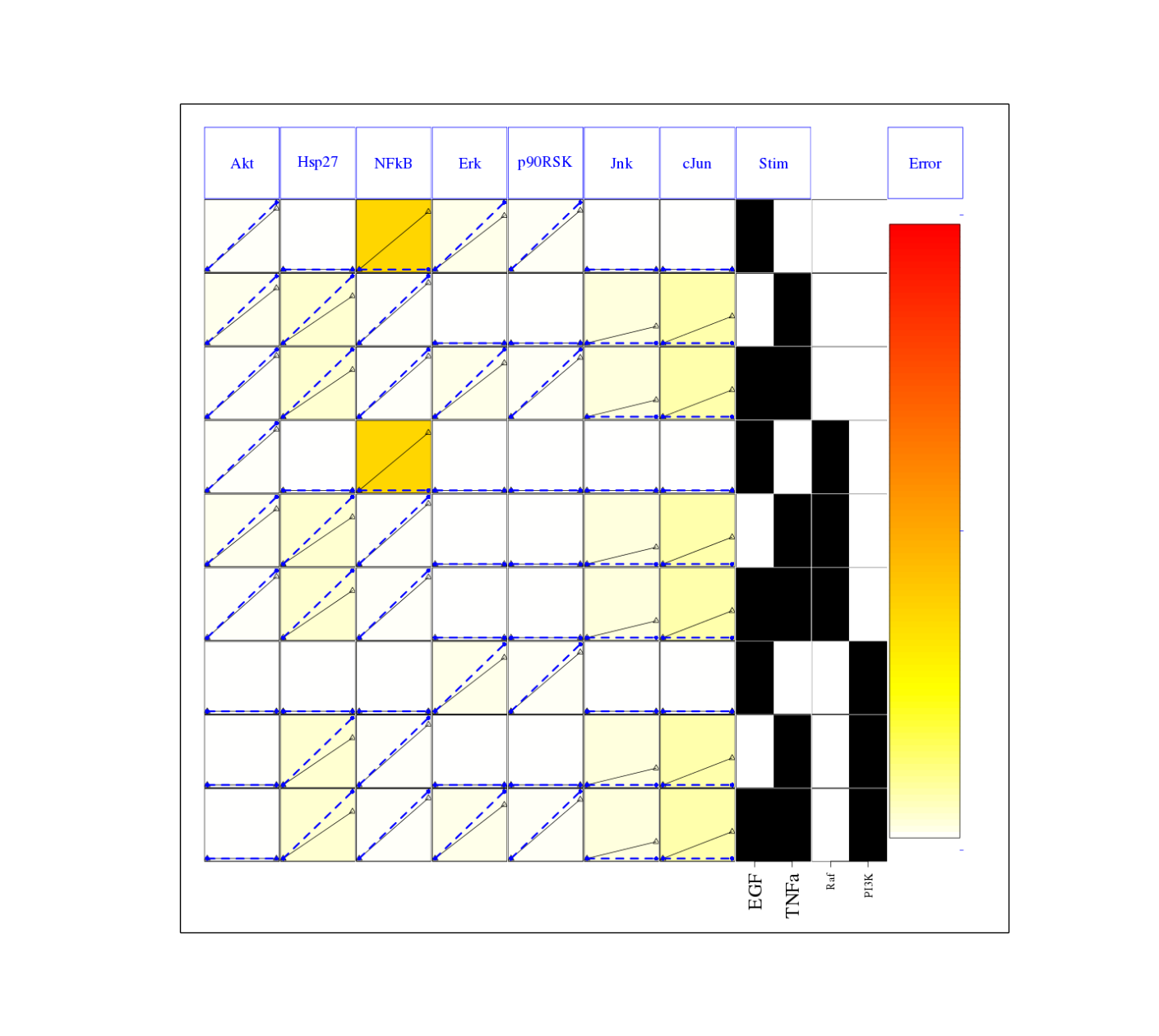}}
\caption{Fitness plot between the data and the best logical model. The plot is
generated by CellNOptR via \textbf{cellnopt.wrapper}. See text for code snippet
and more details. \DUrole{label}{figfit}}
\end{figure}

Note that \textbf{cellnopt.wrapper} is designed to provide a full access to
CellNOptR functionalities only. Yet, for end-users, it is often required to
manipulate the PKN or MIDAS data structures. This was the main motivation to
design \textbf{cellnopt.core} to complement CellNOptR.

\subsection{cellnopt.core%
  \label{cellnopt-core}%
}

\subsubsection{PKN%
  \label{pkn}%
}

The \textbf{cellnopt.core} package provides many tools to manipulate and visualise
networks and MIDAS files. It is implemented in Python and makes use of standard
scientific libraries including Pandas, Matplotlib and NetworkX.\begin{figure}[]\noindent\makebox[\columnwidth][c]{\includegraphics[scale=0.35]{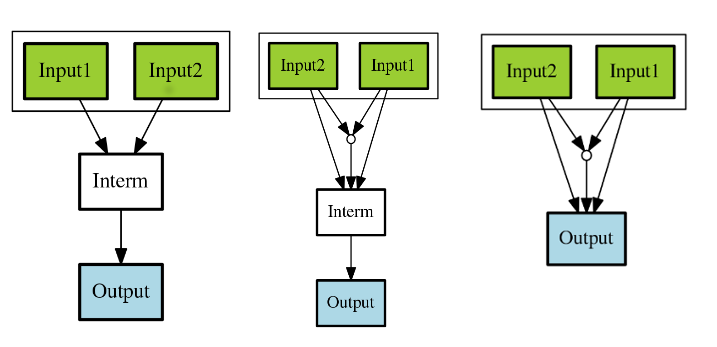}}
\caption{Toy example of a logic model (left panel). Logical and gates are
represented with the small circles (middle).  Logic-based models may also
be compressed so as to simplify the network (right panel). Here the white
node is not required. Removing it does not affect the logic in the network.
\DUrole{label}{fig:cnoproc}}
\end{figure}

Coming back on the simple SIF example shown earlier, we could build it with
the SIF class provided in cellnopt.core but will use another more advanced
structure derived from the directed graph data structure provided by NetworkX.
This class called \textbf{CNOGraph} has dedicated methods to design logic model.
Although you can add nodes and edges using NetworkX methods, you can also add
reactions as follows:\begin{Verbatim}[commandchars=\\\{\},numbers=left,firstnumber=1,stepnumber=1,fontsize=\footnotesize,xleftmargin=2.25mm,numbersep=3pt]
\PY{k+kn}{from} \PY{n+nn}{cellnopt.core} \PY{k+kn}{import} \PY{n}{CNOGraph}
\PY{n}{c}\PY{o}{=} \PY{n}{CNOGraph}\PY{p}{(}\PY{p}{)}
\PY{n}{c}\PY{o}{.}\PY{n}{add\PYZus{}reaction}\PY{p}{(}\PY{l+s}{\PYZdq{}}\PY{l+s}{Input2=Interm}\PY{l+s}{\PYZdq{}}\PY{p}{)}
\PY{n}{c}\PY{o}{.}\PY{n}{add\PYZus{}reaction}\PY{p}{(}\PY{l+s}{\PYZdq{}}\PY{l+s}{Input1=Output}\PY{l+s}{\PYZdq{}}\PY{p}{)}
\PY{n}{c}\PY{o}{.}\PY{n}{add\PYZus{}reaction}\PY{p}{(}\PY{l+s}{\PYZdq{}}\PY{l+s}{Interm=Output}\PY{l+s}{\PYZdq{}}\PY{p}{)}
\PY{n}{c}\PY{o}{.}\PY{n}{\PYZus{}signals} \PY{o}{=} \PY{p}{[}\PY{l+s}{\PYZdq{}}\PY{l+s}{Output}\PY{l+s}{\PYZdq{}}\PY{p}{]}
\PY{n}{c}\PY{o}{.}\PY{n}{\PYZus{}stimuli} \PY{o}{=} \PY{p}{[}\PY{l+s}{\PYZdq{}}\PY{l+s}{Input1}\PY{l+s}{\PYZdq{}}\PY{p}{,} \PY{l+s}{\PYZdq{}}\PY{l+s}{Input2}\PY{l+s}{\PYZdq{}}\PY{p}{]}
\PY{n}{c}\PY{o}{.}\PY{n}{plot}\PY{p}{(}\PY{p}{)}
\end{Verbatim}
where the = sign (A=B) indicates an activation. Inhibitions are encoded
as !A=B, \emph{and} as A\textasciicircum{}B=C and \emph{or} as A+B=C. The
results is shown in Figure \DUrole{ref}{fig:cnoproc} (left panel). By default all nodes
are colored in white but list of stimuli, inhibitors or signals may be provided
manually (line 6,7).

The training of the model to the data may also require to add AND gates, which
is performed as follows:\begin{Verbatim}[commandchars=\\\{\},numbers=left,firstnumber=1,stepnumber=1,fontsize=\footnotesize,xleftmargin=2.25mm,numbersep=3pt]
\PY{n}{c}\PY{o}{.}\PY{n}{expand\PYZus{}and\PYZus{}gates}\PY{p}{(}\PY{p}{)}
\end{Verbatim}
resulting in the model shown in Figure \DUrole{ref}{fig:cnoproc} (middle panel). You
can also compress the network to remove nodes that do not change the logic as
shown in Figure \DUrole{ref}{fig:cnoproc} (right panel):%
\begin{quote}\begin{verbatim}
c.compress()
\end{verbatim}

\end{quote}
On top of the graph data structure, we have also added the split/merge
methods, which can be used to split/merge a protein node into
its variants (e.g., AKT1 and AKT2 instead of just AKT). It can also be used
in the context of mass-spectrometry where measurements of phosphorylation are
made on each peptide individually rather than on the whole protein; number of
peptides varies from a few to dozens of peptides per protein. Consider this
simple example:\begin{Verbatim}[commandchars=\\\{\},numbers=left,firstnumber=1,stepnumber=1,fontsize=\footnotesize,xleftmargin=2.25mm,numbersep=3pt]
\PY{n}{c}\PY{o}{.}\PY{n}{split\PYZus{}node}\PY{p}{(}\PY{l+s}{\PYZdq{}}\PY{l+s}{Interm}\PY{l+s}{\PYZdq{}}\PY{p}{,} \PY{p}{[}\PY{l+s}{\PYZdq{}}\PY{l+s}{Interm1}\PY{l+s}{\PYZdq{}}\PY{p}{,} \PY{l+s}{\PYZdq{}}\PY{l+s}{Interm2}\PY{l+s}{\PYZdq{}}\PY{p}{]}\PY{p}{)}
\PY{n}{c}\PY{o}{.}\PY{n}{plot}\PY{p}{(}\PY{p}{)}
\end{Verbatim}
The split/merge by hand would be tedious on large networks but
is automated with the CNOGraph data structure taking into account AND gates
and input edges (activation/inhibition). Once the PKN is designed, you can
export it into SIF format:\begin{Verbatim}[commandchars=\\\{\},numbers=left,firstnumber=1,stepnumber=1,fontsize=\footnotesize,xleftmargin=2.25mm,numbersep=3pt]
\PY{n}{c}\PY{o}{.}\PY{n}{export2sif}\PY{p}{(}\PY{p}{)}
\end{Verbatim}
You can also export the model into a SBML standard dedicated to logic models
called \textbf{SBMLQual}, which keeps track of the OR and AND logical gates
\cite{CHA13}.\begin{figure}[]\noindent\makebox[\columnwidth][c]{\includegraphics[scale=0.55]{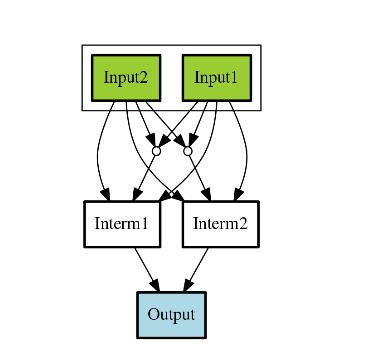}}
\caption{Starting from the middle panel of figure \DUrole{ref}{fig:cnoproc}, CNOGraph data
structure provides a method to split a node into several nodes (updating
AND gates and edges automatically).}
\end{figure}

\subsubsection{DATA%
  \label{id18}%
}

We discussed the MIDAS file format in Figure \DUrole{ref}{figmidas}. CellNOptR
provides tools to look at these data but \textbf{cellnopt.core}
together with Pandas and Matplotlib gives more possiblities. Here is the code
snippet to generate the Figure \DUrole{ref}{figmidas}:\begin{Verbatim}[commandchars=\\\{\},numbers=left,firstnumber=1,stepnumber=1,fontsize=\footnotesize,xleftmargin=2.25mm,numbersep=3pt]
\PY{k+kn}{from} \PY{n+nn}{cellnopt.core} \PY{k+kn}{import} \PY{o}{*}
\PY{n}{m} \PY{o}{=} \PY{n}{XMIDAS}\PY{p}{(}\PY{l+s}{\PYZdq{}}\PY{l+s}{MD\PYZhy{}ToyPB.csv}\PY{l+s}{\PYZdq{}}\PY{p}{)}
\PY{n}{m}\PY{o}{.}\PY{n}{plot}\PY{p}{(}\PY{p}{)}
\end{Verbatim}
The \textbf{XMIDAS} data structure contains 2 dataframes. The first one stores the
experiments. It is a standard dataframe where each row is an experiment and each
column is either a stimuli or an inhibitor. The second dataframe stores the
measurements within a multi-index dataframe where the first dimension is the
cell type, the second is the experiment name, and third is the time point. Each
column corresponds to a protein. The following command shows the time-series of
all proteins in the experiment labelled \textquotedbl{}experiment\_0\textquotedbl{} (no stimuli, no
inhibitors) as shown in Figure \DUrole{ref}{midascut}:\begin{Verbatim}[commandchars=\\\{\},numbers=left,firstnumber=1,stepnumber=1,fontsize=\footnotesize,xleftmargin=2.25mm,numbersep=3pt]
\PY{o}{\PYZgt{}\PYZgt{}}\PY{o}{\PYZgt{}} \PY{n}{m}\PY{o}{.}\PY{n}{df}\PY{o}{.}\PY{n}{ix}\PY{p}{[}\PY{l+s}{\PYZsq{}}\PY{l+s}{Cell}\PY{l+s}{\PYZsq{}}\PY{p}{]}\PY{o}{.}\PY{n}{ix}\PY{p}{[}\PY{l+s}{\PYZsq{}}\PY{l+s}{experiment\PYZus{}0}\PY{l+s}{\PYZsq{}}\PY{p}{]}\PY{o}{.}\PY{n}{plot}\PY{p}{(}\PY{p}{)}
\PY{o}{\PYZgt{}\PYZgt{}}\PY{o}{\PYZgt{}} \PY{n}{m}\PY{o}{.}\PY{n}{experiments}\PY{o}{.}\PY{n}{ix}\PY{p}{[}\PY{l+s}{\PYZsq{}}\PY{l+s}{experiment\PYZus{}0}\PY{l+s}{\PYZsq{}}\PY{p}{]}
\PY{n}{egf}       \PY{l+m+mi}{0}
\PY{n}{tnfa}      \PY{l+m+mi}{0}
\PY{n}{pi3k}\PY{p}{:}\PY{n}{i}    \PY{l+m+mi}{0}
\PY{n}{raf1}\PY{p}{:}\PY{n}{i}    \PY{l+m+mi}{0}
\PY{n}{Name}\PY{p}{:} \PY{n}{experiment\PYZus{}0}\PY{p}{,} \PY{n}{dtype}\PY{p}{:} \PY{n}{int64}
\end{Verbatim}
\begin{figure}[]\noindent\makebox[\columnwidth][c]{\includegraphics[width=\columnwidth]{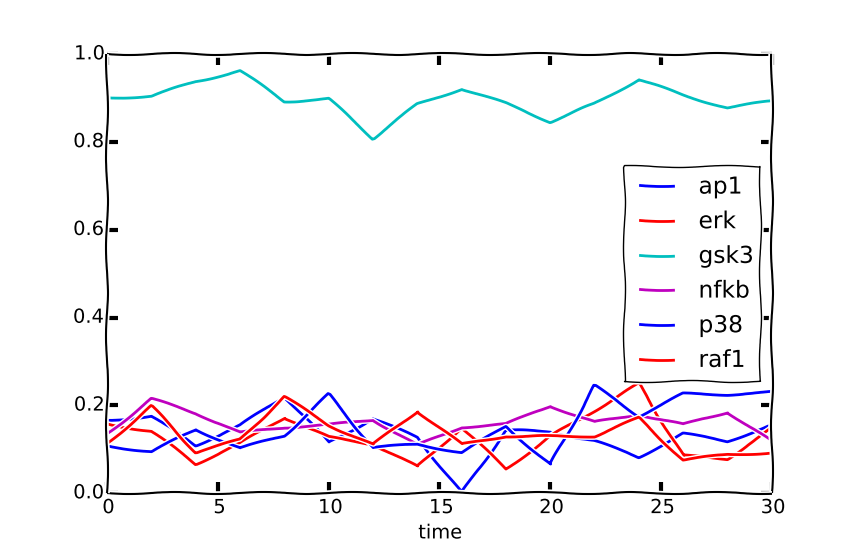}}
\caption{Example of time courses for a given combination of stimuli
and inhibitors. This is the superposition of time series
found in one row of Figure \DUrole{ref}{figmidas}.
One protein level (\emph{gsk3}) is active while others are inactive
when there is no stimuli and no inhibition)
\DUrole{label}{midascut}}
\end{figure}One systematic issue when data is acquired is that it is stored in a
non-standard format so additional scripts are required to translate into a
complex data structure (e.g., MIDAS). Instead of rewriting codes, we can think
about the data as a set of measurements defined by the list of stimuli and
inhibitors, a time point and a value. We can then write one single script that
transforms this list of measurements into a common MIDAS data structure. Here is
an example:\begin{Verbatim}[commandchars=\\\{\},fontsize=\footnotesize]
\PY{k+kn}{from} \PY{n+nn}{cellnopt.core} \PY{k+kn}{import} \PY{n}{MIDASBuilder}
\PY{n}{m} \PY{o}{=} \PY{n}{MIDASBuilder}\PY{p}{(}\PY{p}{)}
\PY{n}{e1} \PY{o}{=} \PY{n}{Measurement}\PY{p}{(}\PY{l+s}{\PYZdq{}}\PY{l+s}{AKT}\PY{l+s}{\PYZdq{}}\PY{p}{,} \PY{l+m+mi}{0}\PY{p}{,} \PY{p}{\PYZob{}}\PY{l+s}{\PYZdq{}}\PY{l+s}{EGFR}\PY{l+s}{\PYZdq{}}\PY{p}{:}\PY{l+m+mi}{1}\PY{p}{\PYZcb{}}\PY{p}{,} \PY{p}{\PYZob{}}\PY{l+s}{\PYZdq{}}\PY{l+s}{AKT}\PY{l+s}{\PYZdq{}}\PY{p}{:}\PY{l+m+mi}{0}\PY{p}{\PYZcb{}}\PY{p}{,} \PY{l+m+mf}{0.1}\PY{p}{)}
\PY{n}{e2} \PY{o}{=} \PY{n}{Measurement}\PY{p}{(}\PY{l+s}{\PYZdq{}}\PY{l+s}{AKT}\PY{l+s}{\PYZdq{}}\PY{p}{,} \PY{l+m+mi}{5}\PY{p}{,} \PY{p}{\PYZob{}}\PY{l+s}{\PYZdq{}}\PY{l+s}{EGFR}\PY{l+s}{\PYZdq{}}\PY{p}{:}\PY{l+m+mi}{1}\PY{p}{\PYZcb{}}\PY{p}{,} \PY{p}{\PYZob{}}\PY{l+s}{\PYZdq{}}\PY{l+s}{AKT}\PY{l+s}{\PYZdq{}}\PY{p}{:}\PY{l+m+mi}{0}\PY{p}{\PYZcb{}}\PY{p}{,} \PY{l+m+mf}{0.5}\PY{p}{)}
\PY{n}{e3} \PY{o}{=} \PY{n}{Measurement}\PY{p}{(}\PY{l+s}{\PYZdq{}}\PY{l+s}{AKT}\PY{l+s}{\PYZdq{}}\PY{p}{,}\PY{l+m+mi}{10}\PY{p}{,} \PY{p}{\PYZob{}}\PY{l+s}{\PYZdq{}}\PY{l+s}{EGFR}\PY{l+s}{\PYZdq{}}\PY{p}{:}\PY{l+m+mi}{1}\PY{p}{\PYZcb{}}\PY{p}{,} \PY{p}{\PYZob{}}\PY{l+s}{\PYZdq{}}\PY{l+s}{AKT}\PY{l+s}{\PYZdq{}}\PY{p}{:}\PY{l+m+mi}{0}\PY{p}{\PYZcb{}}\PY{p}{,} \PY{l+m+mf}{0.9}\PY{p}{)}
\PY{n}{e4} \PY{o}{=} \PY{n}{Measurement}\PY{p}{(}\PY{l+s}{\PYZdq{}}\PY{l+s}{AKT}\PY{l+s}{\PYZdq{}}\PY{p}{,} \PY{l+m+mi}{0}\PY{p}{,} \PY{p}{\PYZob{}}\PY{l+s}{\PYZdq{}}\PY{l+s}{EGFR}\PY{l+s}{\PYZdq{}}\PY{p}{:}\PY{l+m+mi}{0}\PY{p}{\PYZcb{}}\PY{p}{,} \PY{p}{\PYZob{}}\PY{l+s}{\PYZdq{}}\PY{l+s}{AKT}\PY{l+s}{\PYZdq{}}\PY{p}{:}\PY{l+m+mi}{0}\PY{p}{\PYZcb{}}\PY{p}{,} \PY{l+m+mf}{0.1}\PY{p}{)}
\PY{n}{e5} \PY{o}{=} \PY{n}{Measurement}\PY{p}{(}\PY{l+s}{\PYZdq{}}\PY{l+s}{AKT}\PY{l+s}{\PYZdq{}}\PY{p}{,} \PY{l+m+mi}{5}\PY{p}{,} \PY{p}{\PYZob{}}\PY{l+s}{\PYZdq{}}\PY{l+s}{EGFR}\PY{l+s}{\PYZdq{}}\PY{p}{:}\PY{l+m+mi}{0}\PY{p}{\PYZcb{}}\PY{p}{,} \PY{p}{\PYZob{}}\PY{l+s}{\PYZdq{}}\PY{l+s}{AKT}\PY{l+s}{\PYZdq{}}\PY{p}{:}\PY{l+m+mi}{0}\PY{p}{\PYZcb{}}\PY{p}{,} \PY{l+m+mf}{0.1}\PY{p}{)}
\PY{n}{e6} \PY{o}{=} \PY{n}{Measurement}\PY{p}{(}\PY{l+s}{\PYZdq{}}\PY{l+s}{AKT}\PY{l+s}{\PYZdq{}}\PY{p}{,}\PY{l+m+mi}{10}\PY{p}{,} \PY{p}{\PYZob{}}\PY{l+s}{\PYZdq{}}\PY{l+s}{EGFR}\PY{l+s}{\PYZdq{}}\PY{p}{:}\PY{l+m+mi}{0}\PY{p}{\PYZcb{}}\PY{p}{,} \PY{p}{\PYZob{}}\PY{l+s}{\PYZdq{}}\PY{l+s}{AKT}\PY{l+s}{\PYZdq{}}\PY{p}{:}\PY{l+m+mi}{0}\PY{p}{\PYZcb{}}\PY{p}{,} \PY{l+m+mf}{0.1}\PY{p}{)}
\PY{k}{for} \PY{n}{e} \PY{o+ow}{in} \PY{p}{[}\PY{n}{e1}\PY{p}{,}\PY{n}{e2}\PY{p}{,}\PY{n}{e3}\PY{p}{,}\PY{n}{e4}\PY{p}{,}\PY{n}{e5}\PY{p}{,}\PY{n}{e6}\PY{p}{]}\PY{p}{:}
\PY{o}{.}\PY{o}{.}\PY{o}{.}     \PY{n}{m}\PY{o}{.}\PY{n}{add\PYZus{}measurement}\PY{p}{(}\PY{n}{e}\PY{p}{)}
\PY{n}{m}\PY{o}{.}\PY{n}{export2midas}\PY{p}{(}\PY{l+s}{\PYZdq{}}\PY{l+s}{test.csv}\PY{l+s}{\PYZdq{}}\PY{p}{)}
\PY{n}{m}\PY{o}{.}\PY{n}{xmidas}\PY{o}{.}\PY{n}{plot}\PY{p}{(}\PY{p}{)}
\end{Verbatim}
There are many more functionalities available in \textbf{cellnopt.core} especially
to visualise the networks by adding attribute on the edges or nodes, described
within the online documentation.

\subsection{Discussion and future directions%
  \label{discussion-and-future-directions}%
}

In order to call the CellNOptR functionalities within Python, we
decided to use RPy2. There are 16,000 lines of R code in CellNOptR and 4,000
lines of C code, that could not be re-used within Python without being altered.
However, the C code is called by the R functions and therefore does not need any
wrapping functions. Even though the wrapping could be written following RPy2
documentation, however, we had to take into account some considerations. First,
we did not want to  rewrite the documentation. The simplest solution we found
was to implement a \emph{decorator} (called \emph{Rsetdoc}) that appends the R
documentation to the python docstring. Another issue is that it is
non-trivial for the end-user to figure out where to access to the R objects
inside the python function. Consequently, we wrote another decorator
(\emph{Rnames2attributes}) that transforms the R objects into read-only attribute.
So, our wrapping could be as simple as:\begin{Verbatim}[commandchars=\\\{\},fontsize=\footnotesize]
\PY{n+nd}{@Rsetdoc}
\PY{n+nd}{@Rnames2attributes}
\PY{k}{def} \PY{n+nf}{readSIF}\PY{p}{(}\PY{n}{filename}\PY{p}{)}\PY{p}{:}
    \PY{k}{return} \PY{n}{rpack\PYZus{}CNOR}\PY{o}{.}\PY{n}{readSIF}\PY{p}{(}\PY{n}{filename}\PY{p}{)}
\end{Verbatim}
With a straightforward usage, especially for those familiar with the R
commands (same function name):\begin{Verbatim}[commandchars=\\\{\},fontsize=\footnotesize]
\PY{k+kn}{from} \PY{n+nn}{cellnopt.wrapper} \PY{k+kn}{import} \PY{n}{readSIF}
\PY{n}{s} \PY{o}{=} \PY{n}{readSIF}\PY{p}{(}\PY{n}{cnodata}\PY{p}{(}\PY{l+s}{\PYZdq{}}\PY{l+s}{PKN\PYZhy{}ToyMMB.sif}\PY{l+s}{\PYZdq{}}\PY{p}{)}\PY{p}{)}
\PY{n}{s}\PY{o}{.}\PY{n}{interMat}
\PY{o}{\PYZlt{}}\PY{n}{Matrix} \PY{o}{\PYZhy{}} \PY{n}{Python}\PY{p}{:}\PY{l+m+mh}{0x6c0a9e0} \PY{o}{/} \PY{n}{R}\PY{p}{:}\PY{l+m+mh}{0x68f7740}\PY{o}{\PYZgt{}}
\PY{p}{[}\PY{o}{\PYZhy{}}\PY{l+m+mf}{1.000000}\PY{p}{,} \PY{l+m+mf}{0.000000}\PY{p}{,} \PY{l+m+mf}{0.000000}\PY{p}{,} \PY{o}{.}\PY{o}{.}\PY{o}{.}
\end{Verbatim}
Yet, the design and maintenance of the wrapper has a cost. From the development
point of view, we have to keep in mind that the wrapper and the R code have to
be closely managed either by the same developer or team of developers so that
the two codes are maintained and updated synchronously. The second issue is
that a high-level interface such as RPy2 may have a cost on performance. This is
not apparent on a simple script with only a few function calls, but may be
obvious when calling a function a million times (e.g., to perform an
optimisation of a CellNOptR objective function). Although not as elegant,
an alternative to RPy2 is to use the \emph{subprocess} Python module, which could
call a static R pipeline.

\section{BioServices%
  \label{bioservices}%
}

\subsection{Context and motivation%
  \label{context-and-motivation}%
}

In order to construct the PKN required by CellNOpt, we need to access to web resources
such as signalling pathways or protein identifiers. Many resources can be
accessed to in a programmatic way thanks to web services. Building applications
that combine several of them would benefit from a single framework. This was the
main reason to develop \textbf{BioServices}, which is a comprehensive Python
framework that provides programmatic access to major bioinformatics web services
(e.g., KEGG, UniProt, BioModels, etc.).

Two protocols are used to access to web services (i) REST (Representational
State Transfer) and (ii) SOAP (Simple Object Access Protocol). REST has an
emphasis on readability and each resource corresponds to a unique URL.
Operations are carried out via standard HTTP methods
(e.g. GET, POST). SOAP uses XML-based messaging protocol to encode request and
response messages using WSDL (Web Services Description Language).

In order to build applications that
integrate several web services, one needs to have expertise in (i) HTTP
requests, (ii) SOAP protocol, (iii) REST
protocol, (iv) XML parsing to consume the XML messages and
(v) related bioinformatics fields. Consequently, the composition of workflows
or design of external applications based on several web services can be
challenging. BioServices hides the technical aspects of accessing to web
services thereby giving access to a service in a few lines of codes.

\subsection{Approach and Implementation%
  \label{approach-and-implementation}%
}

For developers, there is a class dedicated to REST protocol, and a class
dedicated to WSDL/SOAP protocol. With these classes in place, it is then
straightforward to create a class dedicated to new web service given its URL.
Let us consider WikiPathway \cite{WP09}, which uses a WSDL protocol:\begin{Verbatim}[commandchars=\\\{\},numbers=left,firstnumber=1,stepnumber=1,fontsize=\footnotesize,xleftmargin=2.25mm,numbersep=3pt]
\PY{k+kn}{from} \PY{n+nn}{bioservices} \PY{k+kn}{import} \PY{n}{WSDLService}
\PY{n}{url} \PY{o}{=}\PY{l+s}{\PYZdq{}}\PY{l+s}{http://www.wikipathways.org/}\PY{l+s}{\PYZdq{}}
\PY{n}{url} \PY{o}{+}\PY{o}{=} \PY{l+s}{\PYZdq{}}\PY{l+s}{wpi/webservice/webservice.php?wsdl}\PY{l+s}{\PYZdq{}}
\PY{k}{class} \PY{n+nc}{WikiPath}\PY{p}{(}\PY{n}{WSDLService}\PY{p}{)}\PY{p}{:}
   \PY{k}{def} \PY{n+nf}{\PYZus{}\PYZus{}init\PYZus{}\PYZus{}}\PY{p}{(}\PY{n+nb+bp}{self}\PY{p}{)}\PY{p}{:}
     \PY{n+nb}{super}\PY{p}{(}\PY{n}{WikiPath}\PY{p}{,} \PY{n+nb+bp}{self}\PY{p}{)}\PY{o}{.}\PY{n}{\PYZus{}\PYZus{}init\PYZus{}\PYZus{}}\PY{p}{(}\PY{l+s}{\PYZdq{}}\PY{l+s}{WP}\PY{l+s}{\PYZdq{}}\PY{p}{,} \PY{n}{url}\PY{o}{=}\PY{n}{url}\PY{p}{)}
\PY{n}{wp} \PY{o}{=} \PY{n}{WikiPath}\PY{p}{(}\PY{p}{)}
\PY{n}{wp}\PY{o}{.}\PY{n}{methods} \PY{c}{\PYZsh{} or wp.serv.methods}
\end{Verbatim}
All public methods are shown in the \emph{wp.methods} attribute. A developer can
then access directly to those methods or wrap them to add robustness, quality
and documentation. Let us now use this service to obtain a list of signalling
pathways that contains the protein \emph{MTOR}:\begin{Verbatim}[commandchars=\\\{\},numbers=left,firstnumber=1,stepnumber=1,fontsize=\footnotesize,xleftmargin=2.25mm,numbersep=3pt]
\PY{k+kn}{from} \PY{n+nn}{bioservices} \PY{k+kn}{import} \PY{n}{WikiPathway}
\PY{n}{s} \PY{o}{=} \PY{n}{WikiPathway}\PY{p}{(}\PY{p}{)}
\PY{n}{pathways} \PY{o}{=} \PY{n}{s}\PY{o}{.}\PY{n}{findPathwaysByText}\PY{p}{(}\PY{l+s}{\PYZdq{}}\PY{l+s}{MTOR}\PY{l+s}{\PYZdq{}}\PY{p}{)}
\end{Verbatim}
We can then retrieve a particular signalling pathway and look at it (see Figure
\DUrole{ref}{figwiki}) to  complete our prior knowledge:\begin{Verbatim}[commandchars=\\\{\},numbers=left,firstnumber=1,stepnumber=1,fontsize=\footnotesize,xleftmargin=2.25mm,numbersep=3pt]
\PY{c}{\PYZsh{} Get a SVG representation of the pathway}
\PY{n}{image} \PY{o}{=} \PY{n}{w}\PY{o}{.}\PY{n}{getColoredPathway}\PY{p}{(}\PY{l+s}{\PYZdq{}}\PY{l+s}{WP2320}\PY{l+s}{\PYZdq{}}\PY{p}{)}
\end{Verbatim}
\begin{figure}[]\noindent\makebox[\columnwidth][c]{\includegraphics[scale=0.50]{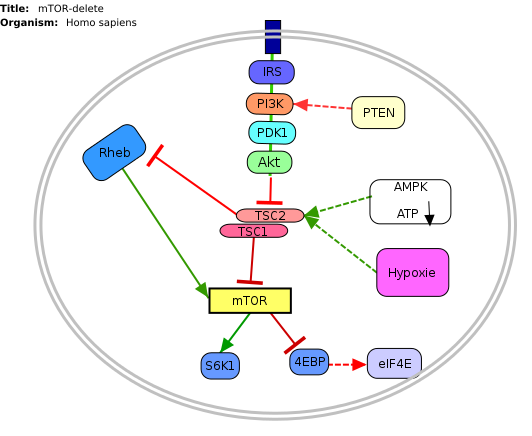}}
\caption{Image obtained from WikiPathway showing a signalling pathway that contains the mTOR protein.
\DUrole{label}{figwiki}}
\end{figure}

\subsection{Combining BioServices with standard scientific tools%
  \label{combining-bioservices-with-standard-scientific-tools}%
}
In general, BioServices does not depend on scientific librairies such as
Pandas so as to limit its dependencies. However, there are a few experimental
methods with a local  \emph{import} so that Pandas is not required during the
installation. In the next example, we will use one of these experimental
methods. UniProt service \cite{UNI14} is useful in CellNOpt for protein
identification and mapping. Let us use it to extract the sequence length of
those proteins. We will then study its distribution. Assuming you have a list of
valid identifiers, just type:\begin{Verbatim}[commandchars=\\\{\},numbers=left,firstnumber=1,stepnumber=1,fontsize=\footnotesize,xleftmargin=2.25mm,numbersep=3pt]
\PY{c}{\PYZsh{} we assume you have a list of entries.}
\PY{k+kn}{from} \PY{n+nn}{bioservices} \PY{k+kn}{import} \PY{n}{UniProt}
\PY{n}{u} \PY{o}{=} \PY{n}{UniProt}\PY{p}{(}\PY{p}{)}
\PY{n}{u}\PY{o}{.}\PY{n}{get\PYZus{}df}\PY{p}{(}\PY{n}{entries}\PY{p}{)}
\end{Verbatim}
Note that the method \emph{get\_df} uses Pandas: it returns a dataframe. One of the
column contains the sequence length. The sequence length distribution can then
be fitted to a SciPy distribution \cite{SCIPY} (using a simple package called
\textbf{fitter},
which is available on PyPi):\begin{Verbatim}[commandchars=\\\{\},numbers=left,firstnumber=1,stepnumber=1,fontsize=\footnotesize,xleftmargin=2.25mm,numbersep=3pt]
\PY{n}{data} \PY{o}{=} \PY{n}{df}\PY{p}{[}\PY{n}{df}\PY{o}{.}\PY{n}{Length}\PY{o}{\PYZlt{}}\PY{l+m+mi}{3000}\PY{p}{]}\PY{o}{.}\PY{n}{Length}
\PY{k+kn}{import} \PY{n+nn}{fitter}
\PY{n}{f} \PY{o}{=} \PY{n}{fitter}\PY{o}{.}\PY{n}{Fitter}\PY{p}{(}\PY{n}{data}\PY{p}{,} \PY{n}{bins}\PY{o}{=}\PY{l+m+mi}{150}\PY{p}{)}
\PY{n}{f}\PY{o}{.}\PY{n}{distributions} \PY{o}{=} \PY{p}{[}\PY{l+s}{\PYZsq{}}\PY{l+s}{lognorm}\PY{l+s}{\PYZsq{}}\PY{p}{,} \PY{l+s}{\PYZsq{}}\PY{l+s}{chi2}\PY{l+s}{\PYZsq{}}\PY{p}{,} \PY{l+s}{\PYZsq{}}\PY{l+s}{rayleigh}\PY{l+s}{\PYZsq{}}\PY{p}{,}
    \PY{l+s}{\PYZsq{}}\PY{l+s}{cauchy}\PY{l+s}{\PYZsq{}}\PY{p}{,} \PY{l+s}{\PYZsq{}}\PY{l+s}{invweibull}\PY{l+s}{\PYZsq{}}
\PY{n}{f}\PY{o}{.}\PY{n}{fit}\PY{p}{(}\PY{p}{)}
\PY{n}{f}\PY{o}{.}\PY{n}{summary}\PY{p}{(}\PY{p}{)}
\end{Verbatim}
In this example, it appears that a log normal distribution is a very good guess
as shown in Figure \DUrole{ref}{fig:uniprot}. Code to get the entries and regenerate
this results is available within BioServices documentation as an IPython
\cite{IPYTHON} notebook.\begin{figure}[]\noindent\makebox[\columnwidth][c]{\includegraphics[scale=0.35]{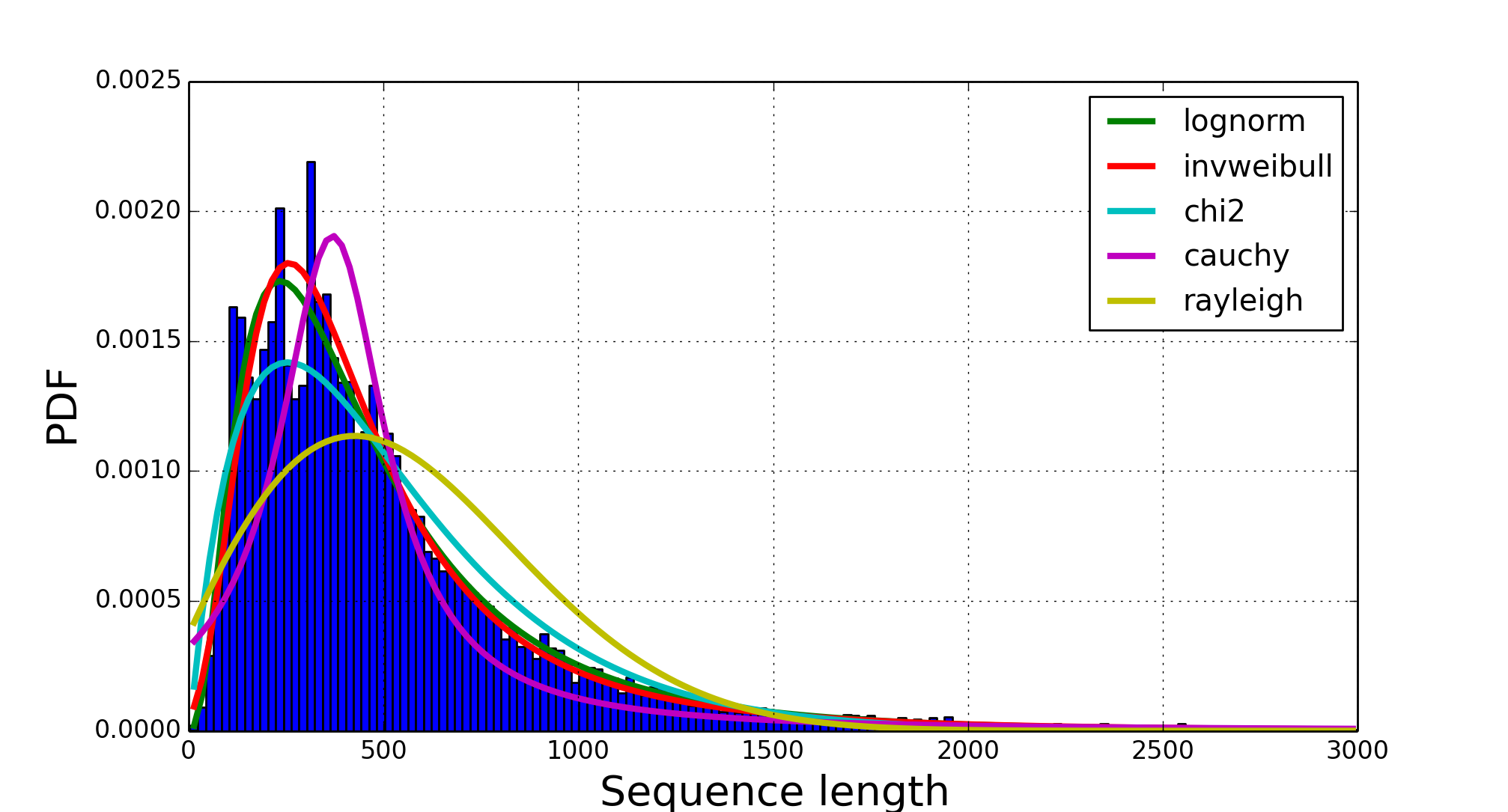}}
\caption{Distribution of the length of 20,000 protein sequence (human).
Distribution was fitted to 80 distributions using SciPy distribution module
and \textbf{fitter} package.
A log normal distribution with parameters fits the length distribution.
See code snippet in the text. \DUrole{label}{fig:uniprot}}
\end{figure}\begin{table*}
\setlength{\DUtablewidth}{0.8\linewidth}
\begin{longtable*}[c]{|p{0.191\DUtablewidth}|p{0.644\DUtablewidth}|}
\hline

REST & 

ArrayExpress, BioMart, ChEMBL, KEGG, HGNC, PDB,
PICR, PSICQUIC, QuickGO, Rhea, UniChem, UniProt,
NCBIBlast, PICR, PSICQUIC \\
\hline

WSDL/SOAP & 

BioModel, ChEBI, EUtils,  Miriam, WikiPathway,
WSDbfetch \\
\hline
\end{longtable*}
\caption{Web services accessible from BioServices (release 1.2.6).
\DUrole{label}{tabbioservices}}\end{table*}

\subsection{Status and future directions%
  \label{status-and-future-directions}%
}

BioServices provides a comprehensive access to bioinformatics web services
within a single Python library. See Table \DUrole{ref}{tabbioservices} for the current
list of services.

The previous example lasts about 20 minutes depending on the network speed.
There are faster way to obtain such information like downloading the database or
flat files. Yet, one need to consider that such files are large (500Mb for
UniProt) and that they may be updated regularly. You may also want to use
several services, which means several flat files. Within a pipeline, you may not
want to provide a set of 500Mb files. In BioServices, the idea is that you do
not necessarily want to download flat files and are willing to wait for the
requests. Future directions of BioServices are two-fold. One is to provide new
web services depending on the user requests and/or contributions. The second
aspect is to make the core functionalities of BioServices faster. This has been
recently achieved with (i) the usage of the \emph{requests} package over the
\emph{urllib2} module (30\% gain) and the buffering or caching of requests to speed up
repetitive requests (also based on the \emph{requests} package).

\section{Conclusions%
  \label{conclusions}%
}

In this paper, we presented \textbf{cellnopt.wrapper} that provides a Python
interface to CellNOptR software. We discussed how and why RPy2 was used to
develop this wrapper. We then presented \textbf{cellnopt.core} that
provides a set of tools to manipulate input data structures required by
CellNOptR (MIDAS and SIF formats amongst others). Visualisation tools are also
provided and the package is linked to Pandas, NetworkX and Matplotlib librairies
making user and developer experience easier and more dynamic. Note that Python
is also used to connect CellNOpt to Answer Set Programming (with the
Caspo package \cite{ASP13}) and to heuristic optimisation methods (\cite{EGE14}).

We also briefly introduced BioServices Python package that allows a
programmatic access to web services used in life sciences. The main interests of
BioServices are (i) to hide technical aspects related to web resource access
(GET/POST requests) so as to foster the integration of new web services (ii) to
put within a single framework many web services.

Source code and extensive on-line documentation are provided on
\url{http://pypi.python.org/pypi} website (bioservices, cellnot.wrapper,
cellnopt.core packages). More information about CellNOptR are available on
\url{http://www.cellnopt.org}.

\section{Acknowledgment%
  \label{acknowledgment}%
}

Authors acknowledge support from EU \emph{BioPreDyn} FP7-KBBE grant 289434.

\end{document}

%% file: page_numbers.tex
\setcounter{page}{35}

%% file: paper.bbl
\begin{thebibliography}{IPYTHON}
\bibitem[ASP13]{ASP13}{

Guziolowski et al.
\emph{Exhaustively characterizing feasible logic models of a signaling network using Answer Set Programming}
Bioinformatics(2013) 29 (18) 2320-2326}
\bibitem[EGE14]{EGE14}{

J. Egea et al.
\emph{MEIGO: an open-source software suite based on metaheuristics for global optimization in systems biology and bioinformatics}
BMC Bioinformatics 2014, 15:136}
\bibitem[UNI14]{UNI14}{

The UniProt Consortium. Nucleic Acids Res. 42: D191-D198 (2014).}
\bibitem[COK13]{COK13}{

T. Cokelaer, D. Pultz, L.M. Harder, J. Serra-Musach and J. Saez-Rodriguez
\emph{BioServices: a common Python package to access biological Web Services programmatically}
Bioinformatics, 29 (24) 3241-3242 (2013)}
\bibitem[WP09]{WP09}{

T. Kelder, AR. Pico, K. Hanspers, MP. van Iersel, C. Evelo, BR. Conklin.
\emph{Mining Biological Pathways Using WikiPathways Web Services.}
PLoS ONE 4(7) (2009). doi:10.1371/journal.pone.0006447}
\bibitem[CNO12]{CNO12}{

C. Terfve, T. Cokelaer, A. MacNamara, D. Henriques, E. Goncalves,
M.K. Morris, M. van Iersel, D.A. Lauffenburger, J Saez-Rodriguez.
\emph{CellNOptR: a flexible toolkit to train protein signaling networks to data using multiple logic formalisms.}
BMC Systems Biology, 2012, 6:133}
\bibitem[CHA13]{CHA13}{

C. Chaouiya et al.
\emph{SBML qualitative models: a model representation format and infrastructure to foster interactions between qualitative modelling formalisms and tools}
BMC Systems Biology 2013, 7:135}
\bibitem[IPYTHON]{IPYTHON}{

F. Pérez and B. E. Granger. \emph{IPython: A system for interactive scientific computing.}
Computing in Science \& Engineering, 9(3):21-29, 2007. \url{http://ipython.org/}}
\bibitem[HUN07]{HUN07}{

J. D. Hunter. \emph{Matplotlib: A 2d graphics environment.}
Computing in Science \& Engineering, 9(3):90-95, 2007. \url{http://matplotlib.org}}
\bibitem[SCIPY]{SCIPY}{

E. Jones, T. E. Oliphant, P. Peterson, et al. \emph{SciPy: Open source
scientific tools for Python}, 2001-. \url{http://www.scipy.org}}
\bibitem[MCK10]{MCK10}{

W. McKinney
\emph{Data Structures for Statistical Computing in Python} in
Proceedings of the 9th Python in Science Conference, p 51-56 2010}
\bibitem[MIDAS]{MIDAS}{

J. Saez-Rodriguez, A. Goldsipe, J. Muhlich, L. Alexopoulos, B.
Millard, D. A.   Lauffenburger, P. K. Sorger,
\emph{Flexible Informatics for Linking Experimental Data to Mathematical Models via DataRail}.
Bioinformatics, 24:6, 840-847 (2008).}
\bibitem[SAEZ]{SAEZ}{

J. Saez-Rodriguez et al.
\emph{Discrete logic modelling as a means to link protein signalling networks with functional analysis of mammalian signal transduction}
Mol. Syst. Biol. (2009), 5, 331}
\bibitem[MAC12]{MAC12}{

A. MacNamara, C. Terfve, D. Henriques, B. Petilde\{n\}alver Bernabacute\{e\}, and J. Saez-Rodriguez
\emph{State–time spectrum of signal transduction logic models}
2012 Phys. Biol. 9 045003}
\bibitem[IDE01]{IDE01}{

T. Ideker, T. Galitski, L. Hood. \emph{A new approach to decoding life: systems biology.}
Annual Review of Genomics and Human Genetics. 2001;2:343–372.}
\bibitem[KIT02]{KIT02}{

H. Kitano. \emph{Systems biology: a brief overview.}
Science. 2002;295(5560):1662–1664.}
\bibitem[ARI08]{ARI08}{

A.A. Hagberg, D.A. Schult and P.J. Swart,
\emph{Exploring network structure, dynamics, and function using NetworkX}
in Proceedings of the 7th Python in Science Conference (SciPy2008),
pp. 11–15, (2008)}
\bibitem[HAUPT]{HAUPT}{

S. Haupt, M. Berger, Z. Goldberg, Y. Haupt
\emph{Apoptosis - the p53 network}
Journal of Cell Science, (2003), 116, 4077-4085.}
\end{thebibliography}
